# A Transformation Approach for Collaboration Based Requirement Models

Ahmed Harbouche[1], Mohammed Erradi[2] and Aicha Mokhtari[3]

[1]Department of Computer Science, Hassiba Ben Bouali University, Chlef, Algeria

Ah.harbouche@gmail.com

[2]Networking & Distributed Systems Research Group, SIME Research Lab, TIES Team, ENSIAS, Mohammed V Souissi University, Rabat, Morocco

erradi@ensias.ma

[3]Knowledge representation and reasoning Team, LRIA, Houari Boumedienne University, Algiers, Algeria

Aissani_Mokhtari@yahoo.fr

## Abstract

*Distributed software engineering is widely recognized as a complex task. Among the inherent complexities is the process of obtaining a system design from its global requirement specification. This paper deals with such transformation process and suggests an approach to derive the behavior of a given system components, in the form of distributed Finite State Machines, from the global system requirements, in the form of an augmented UML Activity Diagrams notation. The process of the suggested approach is summarized in three steps: the definition of the appropriate source Meta-Model (requirements Meta-Model), the definition of the target Design Meta-Model and the definition of the rules to govern the transformation during the derivation process. The derivation process transforms the global system requirements described as UML diagram activities (extended with collaborations) to system roles behaviors represented as UML finite state machines. The approach is implemented using Atlas Transformation Language (ATL).*

## 1. Introduction

Distributed software systems development is an increasingly complex task. Usually the global behavior of distributed systems is not achieved by a single component but by a set of collaborative components. Such global behavior can be decomposed into partial behaviors performed by different system components. Therefore an automatic transformation approach is needed to derive the behavior of these components from the global system requirements. The process of deriving the behavior of system components is very important in the development of distributed systems such as information systems, multi-agent systems and distributed applications. Such automatic derivation process may lead to valid and robust systems by avoiding the error prone manual design activities.

In this work we suggest an approach to derive the partial behaviors of a distributed system from its global requirements. We consider that the basic building blocks of the software system behavior are activities that are actually collaborations between several system components or roles. The suggested approach aims to derive the behavior of the system roles by transforming the requirements model, which describes the functional behavior of a system in an abstract way, to the conceptual model, where the various components are represented by their local behavior. The transformation process, governed by a set of rules, will allow the derivation of the local behavior of a component, in the form of a finite state-machine, from the global system requirements described by an extended UML activity diagram. Such process derives also the appropriate coordinating messages between the derived FSMs. The UML activity diagram is extended with the use of collaborations between the roles of the designed software system.

The suggested approach is inspired from the existing techniques related to communications protocols synthesis from the services specifications [1, 7]. It is mainly based on the technique of meta-model transformations. Such approach consists in defining the source and target meta-models, specifying the transformation rules governing the derivation process and the constraints to be preserved between the different levels of the used models. The development process of the proposed approach is based on the following steps:

1. Defining the requirements meta-model to specify the global system requirements of a given system.
2. Deciding of the meta-model to describe the system at the design level that reflects the local behavior of each system role.
3. Defining the rules to govern the models transformations during the derivation process.

The system requirements are described using UML activity diagrams where the core activities are UML collaborations. At the design level the derived roles behaviors are produced as finite state machines.

This paper is organized as follows. Section 2 presents the related works. Section 3 presents the basic meta-models used as source and target meta-models of the transformation process. Section 4 presents the proposed derivation process. Section 5 presents an illustrative case study. Section 6 presents the implementation of the system design derivation process. Section 7 concludes this paper.

## 2. RELATED WORKS

Many research works have been done on models transformation. In this direction, we could mention the works presented in [2], [3] and [4]. In [2] the suggested approach considers system requirements represented by business processes using BPMN notation. Various UML Use Cases which present some aspects of an information system requirements specification are obtained from business processes by using QVT [5] rules. Also an analytic method of transformation from CIM to PIM level of MDA is described in [3].This approach is based on the transformation of business requirements represented by DFD (Data Flow Diagrams) to UML models. The Work addressed in [4] is concerned with addressing model interoperability between sequence diagrams and Petri Nets, through a model transformation. It introduces a model transformation framework which supports transition between these two models. A designer creates a model of a system as Sequence Diagrams using UML tools, and performs the required analysis using Petri Net tools. However, these works did not consider the case of distributed applications where specific properties such as message exchange and collaborations between the distributed components need to be addressed. Lucena and al. present the STREAM (A Strategy for Transition between Requirements Models and Architectural Models) in [6]. The goal of STREAM is to generate an architectural Model described in Acme architectural description language from i* requirements model.

In fact, the global behavior of distributed systems is not achieved by a single component but by a set of collaborative components. Such behavior can be decomposed into a set of partial behaviors performed by different system components. Therefore a transformation approach is needed to automatically derive the behavior of these components from the overall specification of the global system requirements. The process of deriving the behavior of components of the system is very important in the development of distributed systems. Bochmann [7] suggests an algorithm for the derivation of behavior based on the behavior expressions of collaborations and sub-collaborations as basic constructs for the global system requirements.

In this work we adopt Bochmann's transformation algorithm which transforms collaboration expressions to obtain roles behavior expressions. In our approach we get inspired from such transformation algorithm to provide a derivation process based on meta-models transformation. Our approach starts with defining the source meta-model of UML activity diagrams extended with collaborations, and the target meta- model of UML state machines. Such extended UML activity diagrams are used to describe the system requirements, and allow representing concepts that could not be described using expressions, as done in [7]. This is the case of the conditions (guards) embedded in the choice structures and the repetition structure. In addition, our transformation approach considers also all control nodes specified in the UML activity diagram such as Merge Node and Join Node. In our approach, the derived behaviors of the system roles are UML state machines which could be used for an automatic generation of code.

# 3. THE BASIC META-MODELS

## 3.1. The requirements meta-model

Activity diagram is the only OMG standard notation for modeling Business Process (BPs) and workflow [8]. AD is intended to describe a sequence of actions that may belong to several objects. In order to represent the system requirements, we use UML Activity Diagram (AD) extended with collaborations. The collaborations are used as the main blocks of activities for building requirements models [9]. They define the partial behavior of the entities, called roles, and they give a precise definition of the service behavior using interaction diagrams, activity diagrams or state machines. UML Activity diagrams are suitable to represent choreography of collaborations and sub-collaborations which are the basic activities in a composite collaboration describing the global system requirements. They can express sequential behavior, alternative behavior, competing behavior (parallel composition), repetitive behavior, as well as interruptions.

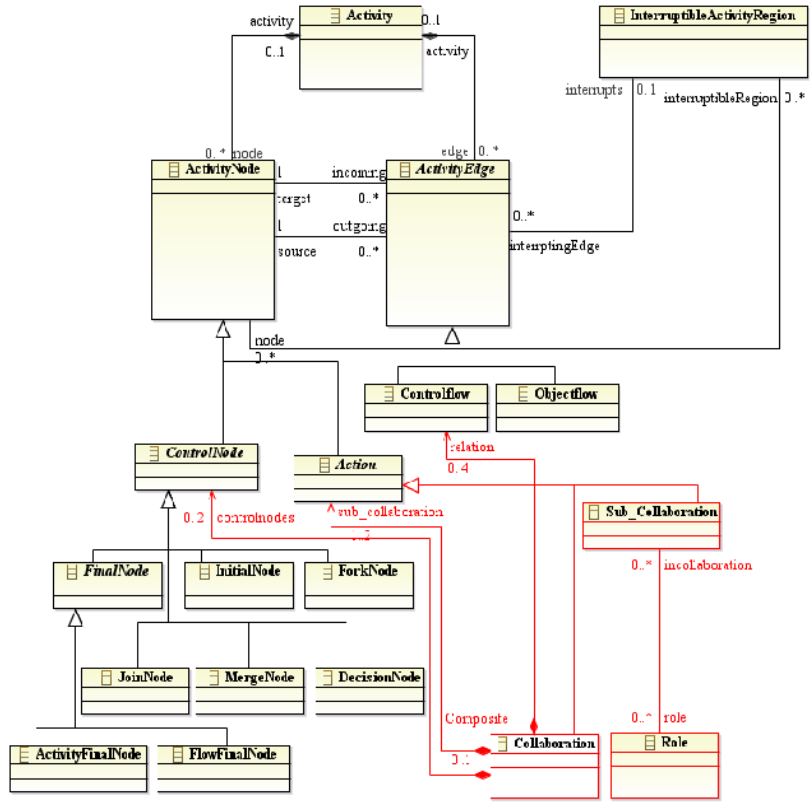

Figure 1. The requirements meta-model.

The proposed requirements meta-model (Figure 1) is considered as the source meta-model of the transformations. It defines the meta-model activity diagram with its main classes and associations while considering collaborations as the basic activities. An activity diagram consists in several *ActivityNode* and *ActivityEdge*. An *ActivityEdge* class allows specifying the control flow (connections) between classes *ActivityNode*. The class *ActivityNode* specifies the different sequencing operators of the activities defined by the class *ControlNode*. This class also defines activities and interruptions. The activities at the meta-model level are defined as "Collaborations" involving multiple roles. A collaboration may consist in one or two collaborations or sub-collaborations. At the models level, the global system requirements are defined as a composition of activity diagrams and collaborations; and considered as an instance of the requirements meta-model.

### 3.2. The target meta-model

The behavior of the various components or roles of a system could be modeled using UML activity diagrams or finite state machines. The transformations of these models to code could be performed and automated. Producing code is out of the scope of this work.

The UML state machine meta-model [10] is used as the target meta-model of the transformation process. A finite state machine diagram allows describing the behavior of a system role. The main classes in this meta-model are the *State* and *Transition* classes. The *State* class models a situation or a significant step in the life cycle of a system component. The *Transition* class models a relationship between a source and a target states, while representing how the component can respond to the occurrence of a given event when the component is in the source state. At the models level, the behavior of a system role is described by a state machine resulting from the derivation process; as an instance of the target meta-model.

## 4. THE DERIVATION PROCESS

### 4.1. The models transformation based derivation process

The derivation process of our approach consists in establishing the transformations from the requirements model into the finite state machine model. The definition of the transformation rules, to govern the derivation process, requires the identification of various relationships between the source and the target meta-models. A transformation rule consists in transforming a concept outlined in the source meta-model to a corresponding one in the target meta-model. For this purpose we define the function named *Transform (Requirement_Concept, FSM_Concept, Messages)*. This function performs two types of the requirement concepts transformation: the first type consists in a direct mapping from a requirement concept to an FSM concept (as in the case of control flows and control nodes). The second type consists in transforming the collaboration concept to a composite state. If the collaboration type is a sub-collaboration, This composite state will hold the actions of the concerned role after transformation otherwise the function *Transform* will trigger the transformation of the collaboration. The *Messages* parameter in this function represents the coordination messages that will be included in the generated composite State. The corresponding rules will be described in section 4.2.

The Derivation Process performs, for a given role, the conformity of the source model (Ms) with the requirements meta-model (MMs). Note that the source model represents the global requirements specification of a given system. The process will then determines the appropriate rule to be applied for each concept or construct represented in Ms and it generates a composite state in the target model. This composite state may hide an FSM which is the specification of the actions of such given role. In addition to the actions role, such composite state may also contain the corresponding coordination messages. The derivation process will generate the needed transitions to connect the generated composite state to its predecessor and successor states.

### 4.2. Transformation rules

To perform the derivation, we identify several cases of choreography expressed in the requirements meta-model (as in [7]): sequential behavior, alternative (choice composition), competition (parallel composition), repetition as well as interruptions.

At the level of collaboration, we distinguish the starting roles and terminating roles [7]. A *starting role* is a role that accomplishes an initial action in a collaboration or in one of its sub-collaborations. A *terminating role* is a role that accomplishes a final action in a collaboration or in one of its sub-collaborations (see Figure 2). We consider that the decision is made locally in a given role. Such a role is the starting role in the collaboration choice structure (it should be the starting role in both collaborations of the choice composition). The transformation of the *collaboration* concept triggers the transformation of its sub-collaborations in the case where it is composed. This property requires the definition of recursive transformation rules.

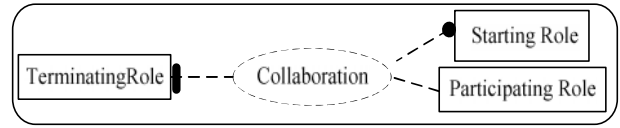

Figure 2. Structure of a collaboration.

Table 1 from [7] shows how to calculate the sets of starting, terminating and participating roles. These sets are calculated for a collaboration depending on the sequencing operators used in the diagram activity.

Two types of sequencing are distinguished [9]: Strong and Weak sequencing. Weak sequencing (Figure 3) specifies that an activity A2 will be executed after another activity A1. While Strong sequencing (Figure 3) implies that all sub-activities of A1 are finished before an activity A2 can begin. Weak sequencing provides only a local order of activities for each system component and does not imply a global order.

Table 1. Rules for calculating the starting, terminating and participating roles [7].

| Choreography Case | Starting roles (SR) | Terminating Roles (TR) | Participating roles (PR) |
|---|---|---|---|
| Sub-collaboration <C> | SR(C) | TR(C) | PR(C)=R |
| Weak Sequencing <C1, C2> | SR(C1)U(SR(C2)-PR(C1)) | TR(C2) U(TR(C1)-PR(C2)) | PR(C1) U PR(C2) |
| Strong Sequencing <C1, C2> | SR(C1) | TR(C2) | PR(C1) U PR(C2) |
| Choice composition <C1, C2> | SR(C1) U SR(C2) = {r} | TR(C1) U TR(C2) | PR(C1) U PR(C2) |
| Strong While loop <C1, C2> | SR(C1) U SR(C2) = {r} | TR(C2) ; SR(C1) if C1=ε | PR(C1) U PR(C2) |
| Weak While loop <C1, C2> | SR(C1) U SR(C2) = {r} | TR(C2) U (TR(C1)-PR(C2) | PR(C1) U PR(C2) |
| Parallel behavior <C1, C2> | SR(C1) U SR(C2) | TR(C1) U TR(C2) | PR(C1) U PR(C2) |

In addition, two coordination messages between different system roles are introduced in the derivation process. We use also the same kind of coordination messages as introduced in Bochmann [7]. Each coordinating message contains the following parameters: a) Source role (Sr), b) Destination role (Dr), c) name of state it-belong to (St). The coordination messages are:

- Flow message for coordinating strong sequencing, named Flowm(Sr, Dr, St).
- Choice indication message for propagating the choice to a role that doesn't participate in the selected alternative in the choice composition structure, named Choicem(Sr, Dr, St).

In the following, we define the transformation rules for the sequential behavior (Strong and weak sequencing) and the choice composition. These rules allow to derive the behavior of different roles involved in a global system requirements specification. Each rule performs the appropriate transformation for a role r participating in a collaboration $C_i$.

### 4.2.1. Strong sequencing between two collaborations

**Rule 1:** If r∈TR(C₁) then Transform(C₁, S₁, Send(Flowm(r, r', S₂)   ∀ r'∈(SR(C₂) - r);

The function transforms the collaboration C₁ to the composite state S₁. This state will hold the actions performed by the role *r* after transformation and includes the action of sending the coordination messages *flowm*. The coordination message *Flowm* is sent by the role *r* to the starting roles of the target collaboration C₂ (except itself if it is a member of this set of roles).

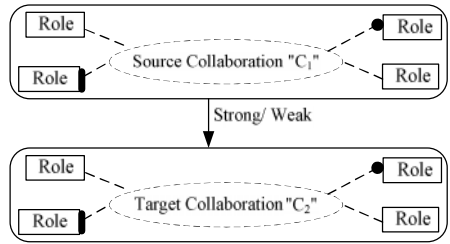

Figure 3. Strong / Weak sequencing.

**Rule 2:** If r∈SR(C₂) then Transform(C₂, S₂, Receive(Flowm(r', r, S₂)  ∀r'∈(TR(C₁) - r);

The collaboration C₂ is transformed to the composite state S₂. This state consists in actions of receiving coordination messages *Flowm* from the terminating roles of the source collaboration C₁ (except from itself) and include the actions performed by *r* after transformation.

**Rule 3:** If r∈((PR(C₁) - TR(C₁)) or (PR(C₂) - SR(C₂)) then Transform(Cᵢ, Sᵢ) ∀ i=1, 2;

**Rule 4:** If r∈(PR(C₁) and PR(C₂)) then Transform(Controlflow, Transition);

This transition connects the states S₁ and S₂ obtained from the transformation of C₁ and C₂.

### 4.2.2. Weak sequencing between two collaborations

The Rule 4 is applied to transform the concept Controlflow.

**Rule 5:** If r∈PR(Cᵢ) then Transform(Cᵢ, Sᵢ)   ∀ i=1, 2 ;

### 4.2.2. Choice between two collaborations (local choice)

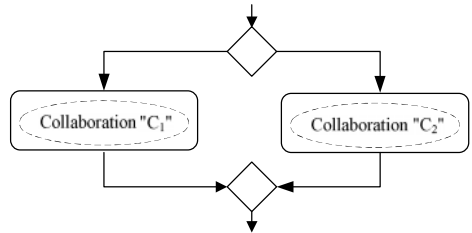

Figure 4. Structure of choice.

**Rule 6:** If r∈SR(Cᵢ) then Transform(Cᵢ, Sᵢ,Send(Choicem(r, r', Sᵢ)
                    ∀ r'∈(PR(Cᵢ·) - PR(Cᵢ)) ∀ i, i'=1,2 and i  i';

**Rule 7:** If r∈PR(Cᵢ) and r∉SR(Cᵢ) then Transform(Cᵢ, Sᵢ)   ∀ i=1, 2;

**Rule 8:** If r∈(PR(Cᵢ) - PR(Cᵢ·))  ∀ i, i'=1,2 and i  i' then

Transform($C_{i'}$, $S_{i'}$, Receive(Choicem($r'$, $r$, $S_{i'}$) $\forall$ $r' \in SR(C_{i'})$ ;

**Rule 9:** If $r \in PR(C_i)$ then Transform (Controlflow, Transition)    $\forall$ i=1, 2;
This transition connects the states $S_1$ and $S_2$ obtained from the transformation of $C_1$ and $C_2$ to the control nodes.

**Rule 10:** If $r \in PR(C_i)$ then Transform(DecisionNode, ChoiceState) $\forall$ i=1, 2;
This state is connected to the composite states $S_1$ and $S_2$ obtained from the transformation of $C_1$ and $C_2$, by the transitions obtained by application of Rule 9.

**Rule 11:** If $r \in PR(C_i)$ then Transform(MergeNode, JunctionState) $\forall$ i=1, 2;
This state is connected to the composite states $S_1$ and $S_2$ obtained from the transformation of $C_1$ and $C_2$, by the transitions obtained by application of Rule 9.

# 4. CASE STUDY

In this section we present a telemedicine case study. Conventionally, when a patient with a stroke is admitted into a hospital (HA), he will be examined by an emergency doctor. The emergency doctor contacts the SAMU (Service for Emergency Medical Assistance) via a regulating doctor to inform him of the emergency admission of a patient who presents symptoms to suspect a stroke (speech disorders, vision disorders, sensor-motor deficits, coordination disorders, etc.). Within the SAMU information system, the regulating doctor creates a medical record based on an initial evaluation using the GCS (Glasgow Coma Score), the hemodynamic status, time of occurrence of signs, time of admission to emergencies, background, anamnesis data and clinical examination, etc…

While remaining in contact with the emergency doctor, the regulating doctor of the SAMU calls the neurologist on duty at CHU (University Hospital Center) and initiates a conference call in order to establish the diagnosis. If appropriate, the patient is urgently transferred to the CHU within an equipped (SMUR) or not-equipped ambulance (VLS) depending on the patient health situation.

In the following we will show the derivation process of the behavior of each role within the global requirements specification of the mentioned case study.

**Step 01:** Specification of the requirements model describing the global system requirements. This model is an instance of the requirements meta-model. The global system requirements is modeled by an activity diagram whose core activities are collaborations and sub-collaborations (Figure 5). The model of the global system requirements is described by a collaboration consisting of two collaborations or sub-collaborations according to the requirements meta-model. This collaboration is a sequence between the *Clinical* sub-collaboration and a composite collaboration named *P-Decision*. The second collaboration *P-Decision* is also a sequence of the *Para-clinical* sub-collaboration and the *Decision* collaboration. The *Decision* collaboration is itself composed of the sub-collaboration *Decision-Making* and the collaboration *During-Transfers*. The last one consists of the choice composition between the collaborations: *Supported by HA* and *Transfer* which is also a choice composition. The collaboration *Transfer* defines a choice between the two sub-collaborations *Sending VLS* and *Sending SMUR*. In the two sub-collaborations, the neurologist (CHU) is a starting and terminating role. The roles patient, emergency doctor (HA) and VLS participate in the two sub-collaborations. While the SMUR participates in the sub-collaboration Sending SMUR and does not participate in Sending VLS.

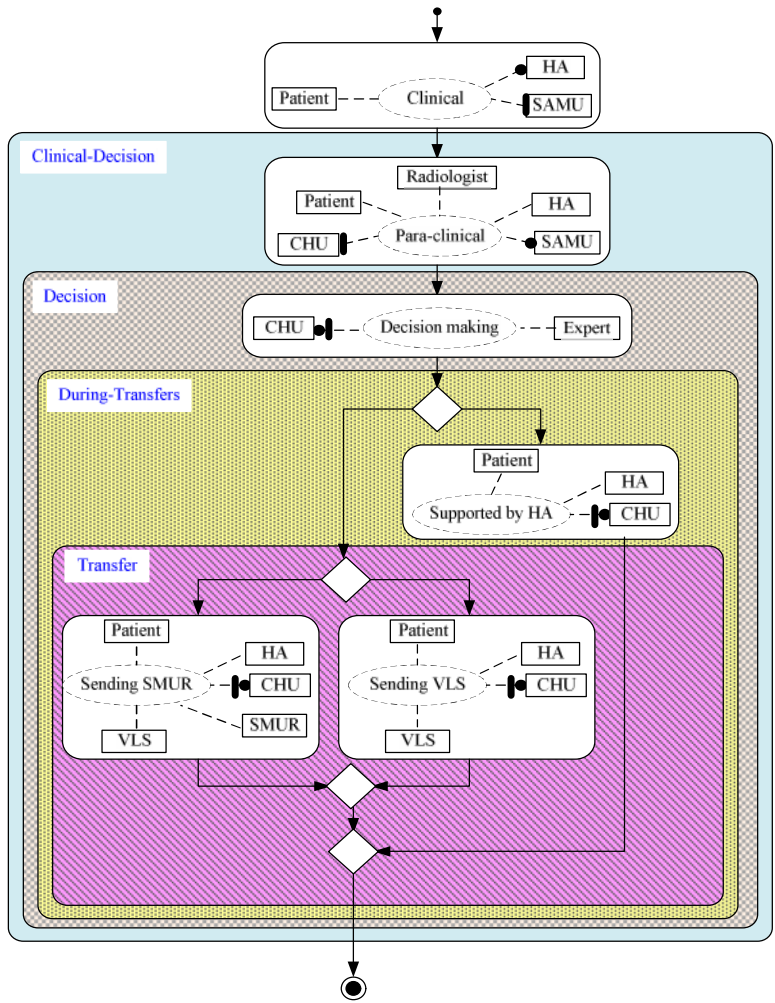

Figure 5. The global system requirements model.

**Step 02:** Applying the model transformation rules that accomplish the derivation of the global system requirements described by the model (Figure 5). These rules are applied to the source model in order to generate a state machine describing each role behavior. We consider in this case the transformation rules that allow deriving the behavior of the neurologist (CHU), the emergency doctor (HA) and the SMUR at the collaboration "Transfer".

The *Transfer* collaboration is a choice composition between the sub-collaborations *Sending VLS* and *Sending SMUR* and the neurologist is responsible for the choice (it is a starting role in the two sub-collaborations). Therefore the derivation process triggers rules 6, 9, 10 and 11. The rule 6 performs the transformation of the actions that it performs at each sub-collaboration (we assume that a sub-collaboration consists in a single action). On the other hand, it must inform the roles not participating in the current collaboration by sending a coordination message *Choicem* for indicating his choice. For example, at the collaboration "Sending VLS", the neurologist sends a coordination message *Choicem* to the role SMUR in order to indicate it that "Sending VLS" was selected. Rules 9, 10 and 11 perform the transformation of Control flow and Control nodes. The derivation process generates a state machine which describes the behavior of the neurologist (CHU) (see Figure 6).

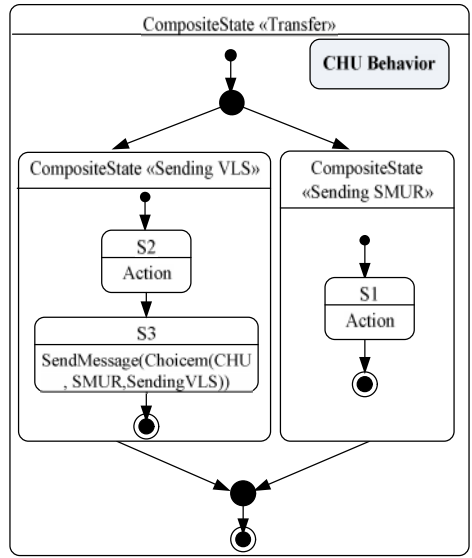

Figure 6. The derived finite state machines describing the CHU behavior.

The emergency doctor (HA) participates in both sub-collaborations *Sending VLS* and *Sending SMUR*. Therefore the derivation process triggers the rules 7, 9, 10 and 11. The rule 7 transforms the actions performed at each sub-collaboration. Rules 9, 10 and 11 accomplish the transformation of Control flow and Control nodes. The emergency doctor (HA) derived behavior is described by the state machine (Figure 7).

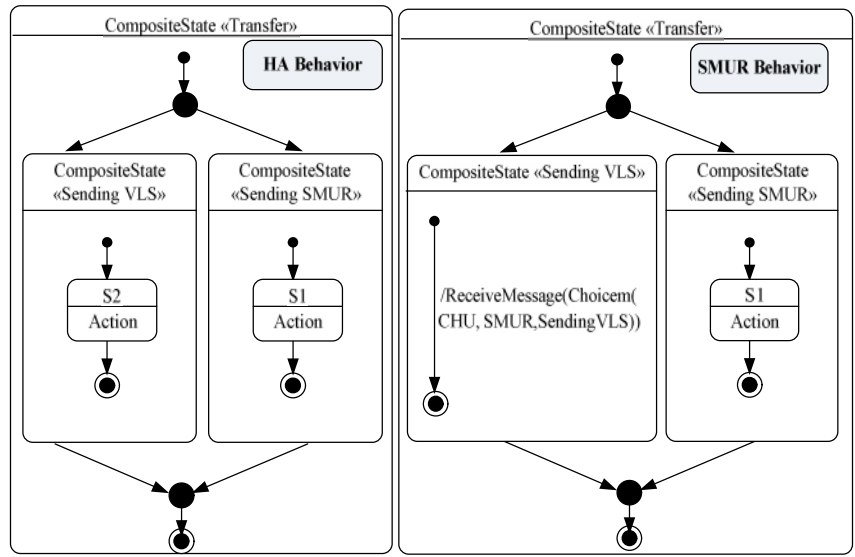

Figure 7. The derived finite state machines describing the HA and SMUR behaviors.

The SMUR participates in the sub-collaboration *Sending SMUR* and not in the *Sending VLS*. The derivation process triggers the rules 7, 8, 9, 10 and 11. The rule 7 accomplish the transformation of the actions performed by the role as a participant at the sub-collaboration *Sending SMUR*. The rule 8 allows the reception of the coordination message *Choicem* sent by the neurologist in case of the selection of the sub-collaboration *Sending VLS*. Rules 9, 10 and 11 perform the transformation of Control flow and Control nodes respectively. The SMUR derived behavior is described by the state machine (Figure 7).

## 6. IMPLEMENTATION

Several researches have suggested approaches and languages for models transformation. Among these languages, we find the ATL language (ATLAS Transformation Language) [11] [12] and the QVT language (Query View Transformation) of the OMG [13] [5]. Both languages exhibit a layered architecture and share common characteristics [12]. We have chosen to express the transformation rules of the derivation process in ATL [11].

According to the adopted derivation process, the implementation of this process requires the following steps:

1. The representation of the requirements meta-model described in UML activity diagram extended with collaborations in *Ecore Diagram* Tool which generate an *Ecore file* named S_MM.ecore described in XMI language.
2. The representation of the target meta-model described in UML state machine diagram in *Ecore Diagram* Tool which generate an *Ecore file* named R_MM.ecore described in XMI language..
3. The representation of the global system requirements model instance of the requirements meta-model in an *Ecore file*.
4. Applying the rules of models transformation specified in ATL language to the source model. This process generates an XMI file containing a finite state machine describing the behavior of each system's role. The role model is consistent with the target meta-model.

In the following, we present the code of the rule 6 and 8 at a sub-collaboration. The sets of starting, terminating and participating roles are calculated using ATL helpers invoked in the ATL transformation rules.

```
Rule CollaborationS2StateSPR {
from   s: MMs!Sub_Collaboration( s.Choice_C12() )
to    r: MMt!CompositeState (Name<- s.Name,
                                    outgoing<-s.outgoing,incoming<-s.incoming,states<-Si,
                                    transition<-ti,states <-sc,transition<-t,states<-st,
                                    transition <-tf, states <-Sf ),
      Si: MMt!InitialState(outgoing <- ti),
      ti: MMt!Transition ( source <-Si , target <- sc),
      sc: MMt!State(action <-act, incoming<-ti,outgoing <-t),
      act: MMt!DomainAction (Name<-s.Name),
      t : MMt!Transition ( source <-sc , target <- st),
      st: MMt!State ( incoming <-t ,outgoing <-tf),
      tf : MMt!Transition ( source <-st , target <- Sf),
      Sf:MMt!FinalState (incoming <-tf)
do   {  thisModule.i<-thisModule.i+1;
         sc.Name<-'S'+thisModule.i.toString(); thisModule.i<-thisModule.i+1;
         st.Name<-'S'+thisModule.i.toString(); thisModule.NameCol<-s.Name;
         st.action <-s.GetRolesC2()->collect(x|thisModule.SendChoicem(x));
         thisModule.j<-thisModule.j+1; ti.Name<-'T'+thisModule.j.toString();
         thisModule.j<-thisModule.j+1; t.Name<-'T'+thisModule.j.toString();
         thisModule.j<-thisModule.j+1; tf.Name<-'T'+thisModule.j.toString();}}
```

The rule 6 (*CollaborationS2StateSPR*) is applied only if the helper Choice_C12() is evaluated to true. This helper determines whether a role is a starting role in the sub-collaboration C1.

**helper context** MMs!Action **def** : Choice_C12() : Boolean =
**if self**.SourceIsDN1() **then if self**.IsinPR()**then if self**.IsinInitial() **then** true **else** false **endif**
                                    **else** false
                                    **endif**
**else** false
**endif**;

The core of this rule consists in:

- Generate a composite State (S) in RolderivBeh(r) conform to the target meta-model $MM_t$;
- Name= C.Name;
- Generate (Initial State) in S;
- Achieve the transformation of the sub-collaboration actions;
- Generate the actions of sending the coordination messages in S to the roles not participating in the sub-collaboration C1 but participating in C2. This set of roles is calculated using the helper GetRolesC2();

**helper context** MMs!Action **def**: GetRolesC2() : OrderedSet (MMs!Role) =
**self**.GetSecond().GetPRCol()->asSet()->
               select (r | **not**(**self**.GetFirst().GetPRCol()->includes(r)));

- Generate (Final State) in S;
- Generate the transitions to connect the different states in the composite State S;
- Connect S to its successor and predecessor states.

The rule 8 (*CollaborationS2StateNPR*) is applied only if the helpers ChoiceN_C12() or ChoiceN_C21() are evaluated to true. These helpers determine whether a role is not participating in the sub-collaboration C1 or C2.

**Rule** CollaborationS2StateNPR {
**from**    s: MMs!Sub_Collaboration (s.ChoiceN_C12() **or** s.ChoiceN_C21() )
**to**    r: MMt!CompositeState (Name<- s.Name, outgoing<-s.outgoing,
                             incoming<-s.incoming,states<-Si,   transition<-ti,states <-sc,
                             transition<-t,states<-st, transition <-tf, states <-Sf ),
      Si: MMt!InitialState(outgoing <- ti),
      ti: MMt!Transition ( source <-Si , target <- sc),
      sc: MMt!State(incoming<-ti,outgoing <-tf),
      tf : MMt!Transition ( source <-sc , target <- Sf),
      Sf:MMt!FinalState (incoming <-tf)
**do**  {  **thisModule**.i<-**thisModule**.i+1;
        sc.Name<-'S'+**thisModule**.i.toString();**thisModule**.NameCol<-s.Name;
        sc.action <-s.GetSRCol->collect(x|**thisModule**.ReceiveChoicem(x));
        **thisModule**.j<-**thisModule**.j+1; ti.Name<-'T'+**thisModule**.j.toString();
        **thisModule**.j<-**thisModule**.j+1; tf.Name<-'T'+**thisModule**.j.toString();}}

The core of this rule consists in:

- Generate a composite State (S) in RolderivBeh(r) conform to the target meta-model $MM_t$;
- Name= C.Name;
- Generate (Initial State) in S;
- Generate the action of receiving the coordination message in S from the starting role of the sub-collaboration. The starting role is calculated using the recursive helper GetSRCol();

**helper context** MMs!Action **def** : GetSRCol():OrderedSet (MMs!Role) =

**if self**.oclIsTypeOf(CIM!Sub_Collaboration) **then self**. GetSRColX()
**else self**.GetSRColY()
**endif**;

- Generate (Final State) in S;
- Generate the transitions to connect the different states in the composite state S;
- Connect S to its successor and predecessor states.

Similar rules are used at the collaboration, but in this case the rules call other rules in order to achieve the collaboration transformation. This is due to the composition of collaboration according to the requirements meta-model. This derivation process generates the finite state machine describing the behavior of each system's role as shown in the telemedicine case study (Figure 6 and Figure7).

## 5. CONCLUSION AND PERSPECTIVES

In this paper, we have presented an approach based on models transformation to derive a distributed system components or roles behavior. This approach consists of a derivation process to derive the behavior of individual components from a given global system requirements. The global system requirements is defined by the requirements model which is an instance of the requirements meta-model based on UML activity diagrams extended with collaborations. A set of rules was defined to govern the transformation process. The target system is derived as state machines that represent the individual behavior of each system's role including the messages required for realizing the collaborations and for ensuring the global coordination of all activities among the different system roles. Our approach was implemented using the ATL language "Atlas Transformation Language". A telemedicine case study was used to illustrate the derivation mechanism. We plan to extend the derivation process by defining the transformation rules to govern the derivation process in the cases of competition (parallel composition) and repetition (While loop with strong and weak sequencing), and including the processing of conditions at the structures of choice between two collaborations and the structure of repetition. Another extension of this approach consists in the derivation of the detailed behavior of a sub-collaboration. The detailed behavior of a sub-collaboration will be described as a sequence diagram. The derivation of a sub-collaboration behavior will be achieved by transforming the sequence diagram into a state machine using a similar models transformation mechanism. We also plan to prove the correctness of the derivation process by validating the transformation rules using a formal proof.

**Authors**

Ahmed Harbouche is a assistant professor in Computer Science since 1993 .  He Obtained his Magister in 1993 at Houari Boumedienne University (Algiers, Algeria) in the area of Artificial intelligence. He is member of the research team "models engineering" of the department of computer science at Hassiba ben Bouali University (Chlef Algeria). He is now  a Ph.d Candidate affiliated with Department of Computing Science of Houari Boumedienne University (Algiers, Algeria). His research interests pertain artificial intelligence, multi-agents systems and Software Engineering. 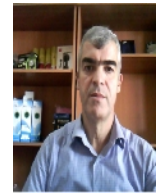

Mohammed Erradi has been a professor in Computer Science since 1986. He has been leading the distributed computing and networking research group since 1994 at ENSIAS (Ecole Nationale d'Informatique et d'Analyse des Systèmes) of Mohammed V-Souissi University (Rabat Morocco). Before joining ENSIAS, Professor Erradi has been affiliated with the University of Sherbrooke and the University of Quebec (UQAT) in Canada. His recent main research interests include Communication Software Engineering, Distributed Collaborative Applications, Security Policies, Reflection and Meta-level Architectures. He obtained his Ph.D. in 1993 at University 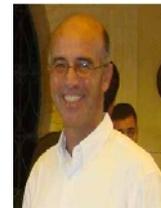 of Montreal in the area of Communicating Software Engineering under the supervision Professor Gregor Von Bochmann. He is leading, at the present time, TENEMO project (A Collaborative Environment for Neuroscience Tele-diagnosis over a Mobile Platform) funded by a French-Moroccan joint research program (2008-2011). He also is currently the Principal Investigator of a number of research projects grants. Professor Erradi has published more than 60 papers in international conferences and journals. He has organized and chaired four international scientific events and has been a member of the program committee in multiple international conferences.